\documentclass[prb,aps,twocolumn,amsmath,amssymb,floatfix,showpacs,
superscriptaddress]{revtex4}

\usepackage[dvips]{graphics}
\usepackage{color}
\definecolor{dred}{rgb}{0,0,0.6}

\begin{document}

\title{\textcolor{dred}{Mobility edge phenomenon in a Hubbard chain: A mean
field study}}

\author{Santanu K. Maiti}

\email{santanu.maiti@isical.ac.in}

\affiliation{Physics and Applied Mathematics Unit, Indian Statistical
Institute, 203 Barrackpore Trunk Road, Kolkata-700 108, India}

\author{Abraham Nitzan}

\affiliation{School of Chemistry, Tel Aviv University, Ramat-Aviv,
Tel Aviv-69978, Israel}

\begin{abstract}

We show that a tight-binding one-dimensional chain composed of interacting 
and non-interacting atomic sites can exhibit multiple mobility edges at 
different values of carrier energy in presence of external electric field. 
Within a mean field Hartree-Fock approximation we numerically calculate 
two-terminal transport by using Green's function formalism. Several 
cases are analyzed depending on the arrangements of interacting and 
non-interacting atoms in the chain. The analysis may be helpful in 
designing mesoscale switching devices.

\end{abstract}


\maketitle

\section{Introduction}

Electronic localization phenomena in one-dimensional ($1$D) quantum 
systems have long been a central problem in condensed matter physics. 
It is well established that in infinite $1$D systems with random site 
potentials, irrespective of the strength of randomness, all the energy 
eigenstates are exponentially localized~\cite{anderson}. Apart from this
Anderson type localization another kind of localization known as 
Wannier-Stark localization is also observed in $1$D materials, even in 
absence of any disorder, when the system is subjected to an external 
electric field~\cite{wan}. For both cases i.e., infinite $1$D systems 
with random site potentials and $1$D chains in presence of 
external electric field, one never encounters {\em mobility edges} 
separating the localized energy eigenstates from the extended ones, since
all eigenstates are localized. However, there are some classes of $1$D 
systems such as, correlated disordered models, quasi-periodic Aubry-Andre 
model where several classic features of mobility edges at some specific 
values of energy are obtained~\cite{dun,sanch,fa,fm,dom,aubry,san6,eco,
das,rolf}. Although the existence of such mobility edges in one- or 
two-dimensional systems has been described by several groups~\cite{eco,das,
rolf,sch,san1,san2}, a comprehensive study of this phenomenon is still 
lacking, particularly in the presence of electron-electron interaction.
Still open fundamental questions are whether some special features exist
in disordered $1$D systems, or in the response of $1$D systems to an 
externally applied electric field even in the presence of electron-electron 
interaction.

In the present article we investigate two-terminal electron transport through 
a $1$D mesoscopic chain composed of interacting and non-interacting atomic 
sites in presence of external electric field. Although some works have been 
done in such superlattice structures~\cite{pai1,pai2,pai3,pai4,wang}, the 
analogous representation of metallic multilayered structures which exhibit 
several novel
features~\cite{multi1,multi2,multi3}, no rigorous effort has been made so 
far, to the best of our knowledge, to unravel the effect of the interplay 
of electron-electron interaction and an imposed external electric field on 
electron transport in such systems. Here we show that a traditional
$1$D lattice with electron-electron interaction, evaluated at the 
Hartree-Fock (HF) mean field (MF) level, is characterized by a mobility 
edge behavior at finite bias voltage. Furthermore, a superlattice structure 
comprising sites on which electron-electron interactions are expressed 
differently (some sites are interacting and some sites are non-interacting) 
is characterized by multiple occurrence of mobility edges at several values 
of the carrier energy. The applicability of mean field approximation in
such superlattice geometries has already been reported in a recent 
work~\cite{meanfield}.

\section{Model and Calculation}

We adopt a tight-binding (TB) framework to describe the model quantum system 
and numerically calculate two-terminal transport within a mean field
\begin{figure}[ht]
{\centering \resizebox*{7.5cm}{2.2cm}{\includegraphics{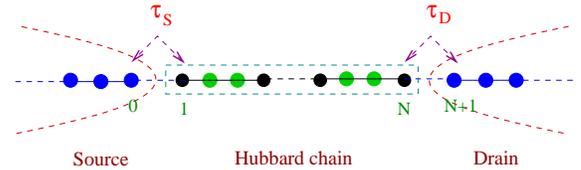}}\par}
\caption{(Color online). A $1$D mesoscopic chain, composed of interacting
(filled black circle) and non-interacting (filled green circle) atomic 
sites, is attached to two semi-infinite $1$D metallic electrodes, 
representing source and drain.}
\label{chain}
\end{figure}
Hartree-Fock approximation using a Green's function formalism. Several 
cases characterized by different arrangements of interacting
and non-interacting atomic sites in the chain, are analyzed. For these 
models we calculate the average density of states (ADOS) and the 
two-terminal transmission probability, and find that sharp crossovers 
from completely opaque to fully or partly transmitting zones take place 
at one or more specific electron energies. This observation suggests the 
possibility of controlling the transmission characteristics by gating the 
transmission zone, and using such superlattice structures as switching 
devices. 

Let us refer to Fig.~\ref{chain} where a $1$D mesoscopic chain, composed 
of non-interacting and interacting atomic sites, is attached to two 
semi-infinite $1$D non-interacting source and drain electrodes. In the 
arrangement of the two different atomic sites shown in Fig.~\ref{chain},
$M$ ($M \ge 1$) non-interacting sites are placed between two interacting 
sites. Here and in what follows we make a restriction that interacting 
atoms are not placed successively. In a Wannier basis, the TB Hamiltonian 
for a $N$-site chain reads,
\begin{eqnarray}
H_C & = &\sum_{i,\sigma}\epsilon_{i\sigma} c_{i\sigma}^{\dagger} 
c_{i\sigma} + \sum_{\langle ij \rangle,\sigma} t \left[
c_{i\sigma}^{\dagger} c_{j\sigma} + c_{j\sigma}^{\dagger} c_{i\sigma} 
\right] \nonumber \\
& + & \sum_i U_i c_{i\uparrow}^{\dagger}c_{i\uparrow} 
c_{i\downarrow}^{\dagger} c_{i\downarrow}
\label{chainham}
\end{eqnarray}
where, $c_{i\sigma}^{\dagger}$ ($c_{i\sigma}$) is the creation 
(annihilation) operator of an electron at the site $i$ with spin $\sigma$
($= \uparrow,\downarrow$), $t$ is the nearest-neighbor hopping element, 
$\epsilon_{i\sigma}$ is the on-site energy of an electron at the site $i$ 
of spin $\sigma$ and $U_i$ is the strength of on-site Coulomb interaction 
where $U_i=0$ for the non-interacting sites. In presence of bias voltage 
$V$ between the two electrodes an electric field is developed and the site 
energies become voltage dependent, 
$\epsilon_{i\sigma}=\epsilon_i^0 + \epsilon_i(V)$,  where $\epsilon_i^0$ 
is a voltage independent term. For the ordered chain $\epsilon_i^0$ is a 
\begin{figure}[ht]
{\centering \resizebox*{6.5cm}{3.5cm}{\includegraphics{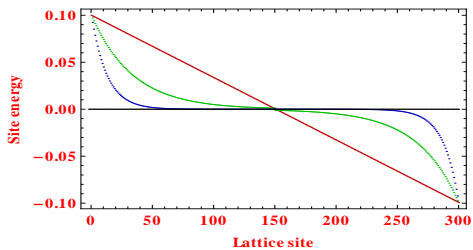}}\par}
\caption{(Color online). Variation of voltage dependent site energies in a
$1$D chain with $300$ lattice sites for three different electrostatic 
potential profiles when the bias voltage $V$ is fixed at $0.2$.}
\label{field}
\end{figure}
constant independent of $i$ that can be chosen zero without loss of
generality, while for the disordered case we select it randomly from a 
uniform ``Box" distribution function in the range $-W/2$ to $W/2$.

The voltage dependence of $\epsilon_i(V)$ reflects the bare electric 
field in the bias junction as well as screening due to longer range e-e
interaction not explicitly accounted for in Eq.~\ref{chainham}. In the
absence of such screening the electric field is uniform along the chain and
$\epsilon_i(V)=V/2-iV/(N+1)$. Below we consider this as well as screened 
electric field profiles, examples of which are shown in Fig.~\ref{field}. 
We will see that the appearance of multiple mobility edges in superlattice
geometries strongly depends on the existence of finite bias and on the 
profile of the bias drop along the chain.

The Hamiltonian for the non-interacting ($U_i=0$) electrodes can be 
expressed as,
\begin{equation}
H_{\mbox{lead}} = \sum_p \epsilon_0 c_p^{\dagger} c_p + \sum_{<pq>} 
t_0 \left(c_p^{\dagger} c_q + c_q^{\dagger} c_p \right)
\label{leadham}
\end{equation}
with site energy and nearest-neighbor intersite coupling $\epsilon_0$ and
$t_0$, respectively. These electrodes are directly coupled to the $1$D chain 
through the lattice sites $1$ and $N$. The hopping integrals between the 
source and chain and between the chain and drain are denoted by $\tau_S$
and $\tau_D$, respectively. 

In the generalized HF approach~\cite{san3,san4,san5,kato,kam}, the full 
Hamiltonian is decoupled into its up-spin and down-spin components by 
replacing the interaction terms by their mean field (MF) counterparts. 
This redefines the on-site energies as 
$\epsilon_{i\uparrow}^{\prime}=\epsilon_{i\uparrow} 
+ U \langle n_{i\downarrow} \rangle$ and
$\epsilon_{i\downarrow}^{\prime}=\epsilon_{i\downarrow} + U \langle 
n_{i\uparrow} \rangle$ where, $n_{i\sigma}=c_{i\sigma}^{\dagger} 
c_{i\sigma}$ is the number operator. With these site energies, the full 
Hamiltonian (Eq.~\ref{chainham}) can be written in the MF approximation
in the decoupled form
\begin{eqnarray}
H_{MF} &=&\sum_i \epsilon_{i\uparrow}^{\prime} 
n_{i\uparrow} + 
\sum_{\langle ij \rangle} t \left[c_{i\uparrow}^{\dagger} c_{j\uparrow} + 
c_{j\uparrow}^{\dagger} c_{i\uparrow}\right] \nonumber \\
& + & \sum_i \epsilon_{i\downarrow}^{\prime} n_{i\downarrow} + \sum_{\langle 
ij \rangle} t \left[c_{i\downarrow}^{\dagger} c_{j\downarrow}
+ c_{j\downarrow}^{\dagger} c_{i\downarrow}\right] \nonumber \\
& - & \sum_i U_i \langle n_{i\uparrow} \rangle \langle n_{i\downarrow} 
\rangle \nonumber \\
&=& H_{C,\uparrow} + H_{C,\downarrow} - 
\sum_i U_i \langle n_{i\uparrow} \rangle \langle n_{i\downarrow} \rangle
\label{equ400} 
\end{eqnarray}
where, $H_{C, \uparrow}$ and $H_{C,\downarrow}$ correspond to the effective 
\begin{figure}[ht]
{\centering \resizebox*{7.5cm}{7cm}
{\includegraphics{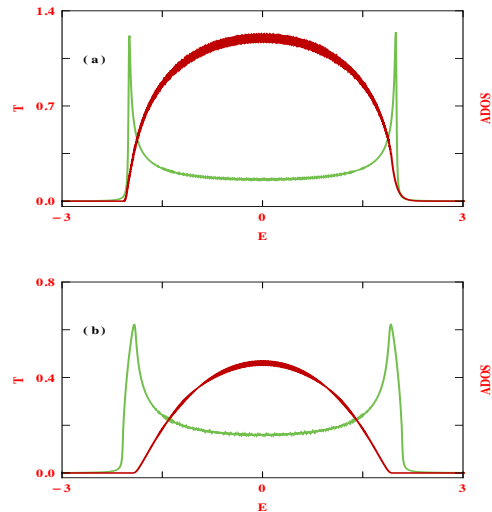}}\par}
\caption{(Color online). Transmission probability $T$ (red color) and 
ADOS (green color) as a function of energy $E$ for a $1$D non-interacting
($U_i=0$ $\forall$ $i$) ordered ($W=0$) chain with $N=300$ sites. The 
electrostatic potential profile varies linearly (red curve in 
Fig.~\ref{field}), with the total potential drop across the chain to be 
(a) $V=0$ and (b) $V=0.2$.}
\label{fullnoninteracting}
\end{figure}
TB Hamiltonians for the up and down spin electrons, respectively. The last 
term provides a shift in the total energy that depends on the mean 
populations of the up and down spin states.
\begin{figure*}[ht]
{\centering \resizebox*{14cm}{10cm}{\includegraphics{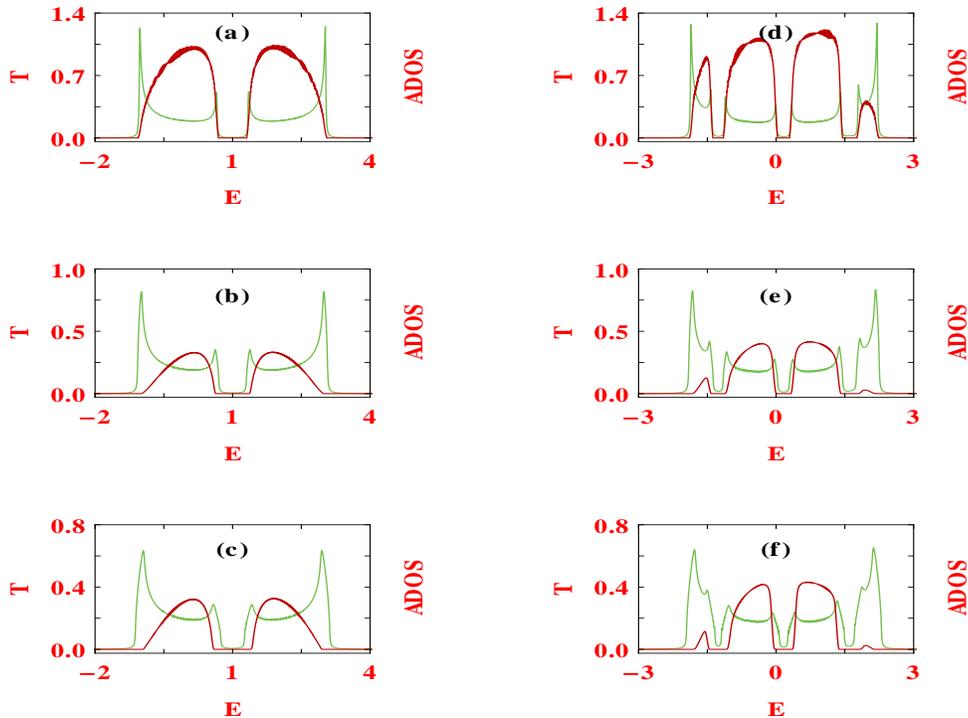}}\par}
\caption{(Color online). Transmission probability $T$ (red color) and ADOS
(green color) as a function of energy $E$ for an ordered ($W=0$) $1$D chain.
The left column corresponds to the case where all sites are interacting 
($U_i=2$), while the right column represents the results for a $1$D 
superlattice geometry where four non-interacting ($U_i=0$) atoms are 
placed between two interacting ($U_i=2$) atoms. The 1st, 2nd and 3rd rows 
correspond to $V=0$, $0.1$ and $0.2$, respectively. All these results are
shown for a linear bias drop along the chain.}
\label{four}
\end{figure*}

With these decoupled Hamiltonians ($H_{C,\uparrow}$ and $H_{C,\downarrow}$) 
of up and down spin electrons, we start our self consistent procedure 
considering initial guess values of $\langle n_{i\uparrow} \rangle$ and 
$\langle n_{i\downarrow} \rangle$. For these initial set of values of 
$\langle n_{i\uparrow} \rangle$ and $\langle n_{i\downarrow} \rangle$, 
we numerically diagonalize the up and down spin Hamiltonians. Then we 
calculate a new set of values of $\langle n_{i\uparrow} \rangle$ and 
$\langle n_{i\downarrow} \rangle$. These steps are repeated until a self 
consistent solution is achieved.

The converged mean field Hamiltonian is a sum of single electron up and
down spin Hamiltonians. The transmission function is therefore a sum
$T(E)=\sum_{\sigma} T_{\sigma}(E)$ where~\cite{datta} 
$T_{\sigma}={\mbox{Tr}} \left[\Gamma_S \, G_{C,\sigma}^r \, 
\Gamma_D \, G_{C,\sigma}^a\right]$. Here, $G_{C,\sigma}^r$ and 
$G_{C,\sigma}^a$ are the retarded and advanced Green's functions, 
respectively, of the chain including the effects of the electrodes. 
$G_{C,\sigma}=\left(E-H_{C,\sigma}-\Sigma_S-\Sigma_D \right)^{-1}$, 
where $\Sigma_S$ and $\Sigma_D$ are the self-energies due to coupling 
of the chain to the source and drain, respectively, while $\Gamma_S$ 
and $\Gamma_D$ are their imaginary parts. 

\section{Results and Discussion}

In what follows we limit ourselves to absolute zero temperature and use 
the units where $c=h=e=1$. For the numerical calculations we choose 
$t=1$, $\epsilon_0=0$, $t_0=3$ and $\tau_S=\tau_D=1$. The energy scale 
is measured in unit of $t$.

Before addressing the central problem i.e., the possibility of getting 
multiple mobility edges in $1$D superlattice geometries, first we explore 
\begin{figure}[ht]
{\centering \resizebox*{6.75cm}{4cm}{\includegraphics{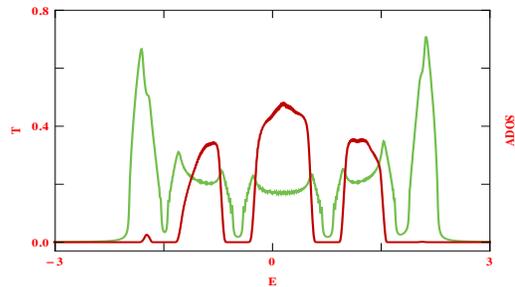}}\par}
\caption{(Color online). Transmission probability $T$ (red color) and 
ADOS (green color) as a function of energy $E$ for a $1$D chain ($N=300$) 
in absence of disorder ($W=0$) with on-site interaction $U_i=2$ and bias
voltage $V=0.2$ that varies linearly along the chain. Here we set $M=5$ i.e,
five non-interacting atoms are placed between two interacting atoms.}
\label{five}
\end{figure}
the effect of finite bias on electron transport in two simple systems, 
one for a standard non-interacting chain and the other for a conventional 
Hubbard chain where all sites are interacting. 

In Fig.~\ref{fullnoninteracting} we show the variation of total transmission
probability (T) together with the average density of states as a function of
energy $E$ for an ordered ($\epsilon_i^0=0$ for all atomic sites $i$ in the 
chain) non-interacting chain for two different magnitudes of the voltage 
bias, assuming a linear bias drop (uniform electric field) across the chain.
In the absence of electric field electron transmission takes place 
throughout the energy band as clearly seen from the spectrum 
Fig.~\ref{fullnoninteracting}(a), since in this case all the energy 
eigenstates are extended. On the other hand, when a finite bias drop takes 
\begin{figure}[ht]
{\centering \resizebox*{6.75cm}{4cm}{\includegraphics{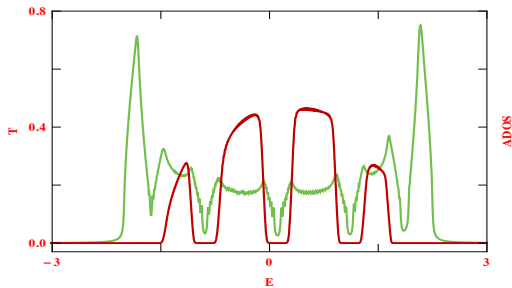}}\par}
\caption{(Color online). Same as Fig.~\ref{five}, with $M=6$.}
\label{six}
\end{figure}
place along the chain, several energy eigenstates appear in the energy 
regions around the band edges for which the transmission probability is
exactly zero (Fig.~\ref{fullnoninteracting}(b)). Therefore, the chain 
appears insulating when Fermi energy is within the zone of zero 
transmission, while finite transmission, $T \ne 0$, is seen more towards 
the band centre. The sharp transition between these regimes illustrates 
the existence of a mobility edge phenomenon under finite bias
\begin{figure}[ht]
{\centering \resizebox*{6.75cm}{4cm}{\includegraphics{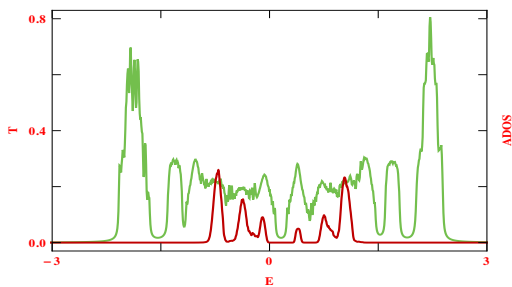}}\par}
\caption{(Color online). Transmission probability $T$ (red color) and 
ADOS (green color) as a function of energy $E$ for a $1$D chain ($N=300$) 
in presence of disorder ($W=0.5$) for the same parameter values used in 
Fig.~\ref{four}(f): $U_i=2$, $V=0.2$ (linear potential profile) and $M=4$.}
\label{fourdis}
\end{figure}
condition. For a finite bias, the localization of energy levels always 
starts from the band edges and the width of the localized energy zones 
can be controlled by the imposed electric field. Obviously, for strong 
enough electric field almost all the energy levels are localized and the 
extended energy regions disappear, so that in this particular case 
metal-insulator (MI) transition will no longer be observed. This 
localization phenomenon in presence of an external electric field has
already been established in the literature, but the central issue of our
present investigation - the interplay between the Hubbard interaction 
strength, the superlattice configuration and the electric field has not 
been addressed earlier.

To explore it, we present in Fig.~\ref{four} the results of a traditional
Hubbard chain where all sites are interacting (left column) together with 
the results of a superlattice geometry where four non-interacting atoms 
are placed between two interacting atoms (right column). The results are 
shown for three different values of the voltage bias, taking a linear bias 
drop along the $1$D chain. For the chain where all sites are interacting a 
single energy gap only appears at the band centre, while in the superlattice 
geometry, depending on the unit cell configuration, multiple energy gaps 
are generated which are clearly visible from the ADOS spectra. Therefore, in 
a superlattice geometry, in presence of external electric field associated
with bias voltage $V$ between two electrodes, zero transmission ($T=0$) 
energy regions exist, and are separated by regions of extended states 
compared to the traditional Hubbard chain, and, it leads to the possibility 
of getting an MI-like transition at multiple energies.

The total number of energy sub-bands in a superlattice geometry for a 
\begin{figure}[ht]
{\centering \resizebox*{6.75cm}{4cm}
{\includegraphics{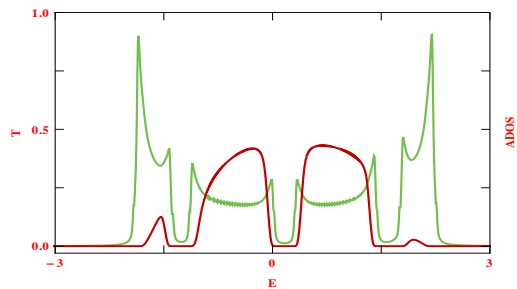}}\par}
\caption{(Color online). Transmission probability $T$ (red color) and 
ADOS (green color) as a function of energy $E$ for a $1$D chain ($N=300$) 
with no disorder ($W=0$). The model parameters are $U=2$, $M=4$ and $V=0.2$, 
with the bias potential profile taken as the green curve given of
Fig.~\ref{field}.}
\label{fourflat1}
\end{figure}
particular energy range generated in the ADOS profile strongly depends 
on structural details i.e., the number $M$ of non-interacting atoms 
between two interacting lattice sites. This is shown in Figs.~\ref{five} 
and \ref{six} which show the ADOS and the transmission probability for 
two models that are identical in all details (see caption to 
Fig.~\ref{five}) except that $M=5$ in Fig.~\ref{five} and $M=6$ in 
Fig.~\ref{six}. These structures show more mobility edge phenomena, that 
is crossovers between fully opaque and a transmitting zone, than in the
corresponding case of Fig.~\ref{four}(f), suggesting a design concept based
on such superlattice structures as a switching devices at multiple energies.

The robustness of the observed behavior can be examined by its sensitivity 
to the presence of disorder. Figure~\ref{fourdis} displays the ADOS spectrum
and the total transmission probability for a $1$D chain in presence of 
diagonal disorder affected by choosing $\epsilon_i^0$ from a uniform
distribution of width $W=0.5$ ($-0.25$ to $+0.5$). An average over $50$
disorder configurations is presented. The resulting ADOS and transmission 
show similar qualitative features, with sharp transitions between localized
and extended spectral regions as seen above for the ordered cases. Note that
the presence of disorder alone can cause state localization. For strong 
enough disorder almost all energy levels get localized, even for such a 
finite size $1$D chain. In this limit such crossover behavior will no longer
be observed. 

\begin{figure}[ht]
{\centering \resizebox*{6.75cm}{4cm}
{\includegraphics{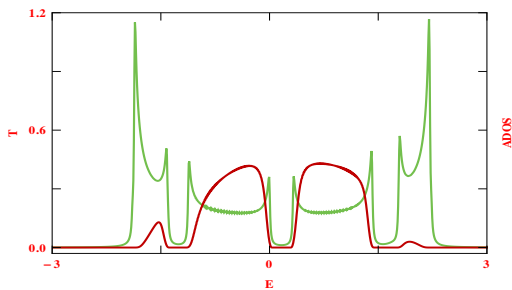}}\par}
\caption{(Color online). Same as Fig.~\ref{fourflat1}, with the electrostatic
potential profile given by the blue curve in Fig.~\ref{field}.}
\label{fourflat2}
\end{figure}

In the calculations presented so far we have assumed a linear drop of the
electrostatic potential along the chain. Figures~\ref{fourflat1} and
\ref{fourflat2} show results obtained for an identical chain length with
other potential profiles that are characteristics of stronger screening.
We see that the localized region gradually decreases with increasing 
flatness of the potential profile in the interior of the conducting bridge.
If the potential drop takes place only at the chain-to-electrode interfaces
i.e., when the potential profile becomes almost flat along the chain the
width of the localized region almost vanishes and the metal-insulator 
transition is not observed, as was the case for the zero bias limit.

Finally, we point out that by locating the Fermi energy in appropriate 
places of the sub-bands, the system can be used as a p-type or an n-type 
semiconductor. For example, let us imagine, at absolute zero temperature, 
the Fermi level is fixed in the localized region which is very close 
to the fully transmitting zone (right hand side). In this case, the 
left sub-bands up to the Fermi level are completely filled with electrons. 
Now, if the energy gap between the Fermi level, pinned in the localized 
region, and the bottom of the transmitting region (right hand side) is 
small enough for electrons to hop, then the system will behave as an 
n-type semiconductor. On the other hand, by reverting the situation we 
can generate a p-type semiconductor where electrons hop from a filled 
transmitting zone (valence band) to unoccupied localized zone (conduction 
band) generating holes in the valence band.

\section{Conclusion}

To summarize, we have investigated in detail the two-terminal finite bias 
electron transport in a $1$D superlattice structure composed of interacting 
and non-interacting atoms. The electron-electron interaction is considered 
in the Hubbard form, and the Hamiltonian is solved within a generalized HF 
scheme. We numerically calculate two-terminal transport by using a Green's 
function formalism and analyze the results for some specific chain 
structures characterized by different arrangements of the atomic sites 
in the chain. Our analysis may be utilized in designing a tailor made 
switching device for multiple values of Fermi energy (or, more practically, 
for different values of a gating potential). The sensitivity of this 
switching action i.e., metal-to-insulator transition and vice versa on 
the electric field variation has also been discussed. Though the results 
presented in this article are worked out at absolute zero temperature limit, 
the results should remain valid even at finite temperatures ($\sim300\,$K) 
since the broadening of the energy levels of the superlattice structure 
due to its coupling with the metal electrodes is much higher than that 
of the thermal broadening~\cite{datta}.

\section{Acknowledgment} 

The research of A.N. is supported by the Israel Science Foundation, the
Israel-US Binational Science Foundation, the European Research Council 
under the European Union's Seventh Framework Program (FP7/2007-2013; ERC
Grant No. 226628) and the Israel-Niedersachsen Research Fund.

\end{document}